\documentclass[10pt,conference,letterpaper]{IEEEtran}
\IEEEoverridecommandlockouts

\usepackage{booktabs}
\usepackage{multirow}

\usepackage{cite}
\usepackage{amsmath,amssymb,amsfonts}
\usepackage{algorithmic}
\usepackage{graphicx}
\usepackage{textcomp}
\usepackage{xcolor}
\def\BibTeX{{\rm B\kern-.05em{\sc i\kern-.025em b}\kern-.08em
    T\kern-.1667em\lower.7ex\hbox{E}\kern-.125emX}}
\usepackage[caption=false,font=footnotesize]{subfig}

\usepackage[hidelinks]{hyperref}

\newif\ifanonymous
\anonymousfalse           

\begin{document}

\title{HoRFFI: High-Openness RF Fingerprint Identification with a Similarity-Enhanced Variational Information Bottleneck}

\ifanonymous
\author{\IEEEauthorblockN{Anonymous Authors}}
\else
\author{
\IEEEauthorblockN{Shuiguang Zeng\textsuperscript{1,2}, Yuxiang Shen\textsuperscript{1,2}, Yuanyu Zhang\textsuperscript{3,*}, Yulong Shen\textsuperscript{3}, Zhiyuan Tan\textsuperscript{4}, and Houbing Herbert Song\textsuperscript{5}}
\IEEEauthorblockA{
\textsuperscript{1}\textit{College of Computer and Cyber Security, Hebei Normal University}, Shijiazhuang, China\\
\textsuperscript{2}\textit{Hebei Key Laboratory of Network and Information Security}, Shijiazhuang, China\\
\textsuperscript{3}\textit{School of Computer Science and Technology, Xidian University},
Xi’an, China\\
\textsuperscript{4}\textit{School of Computing, Engineering and the Built Environment, Edinburgh Napier University},
Edinburgh, U.K.\\
\textsuperscript{5}\textit{Department of Information Systems, University of Maryland, Baltimore County (UMBC)},
Baltimore, USA\\
\textsuperscript{*}Corresponding author:
Yuanyu Zhang (yyuzhang@xidian.edu.cn)}

}
\fi
\maketitle
\begin{abstract}
Radio frequency fingerprint identification (RFFI) is a promising technique for wireless device authentication. However, practical RFFI systems must enroll newly authorized devices while rejecting previously unseen ones, even when the feature extractor is trained on only a few labeled base-device classes, giving rise to a high-openness RFFI problem. Existing open-set recognition methods typically rely on feature spaces learned from a large and diverse set of known-device classes, limiting their applicability in practical scenarios. To address this challenge, we propose HoRFFI, a high-openness RFFI framework that supports scalable device identification and unknown-device rejection using only a small number of labeled training devices. HoRFFI employs a similarity-enhanced variational information bottleneck (SVIB)-based supervision mechanism, which reduces the encoder's dependence on training-class diversity and learns a more transferable embedding space. This supervision mechanism uses feature-space augmentation and clustering to derive inter-sample similarity information, which provides supplementary supervision for regularizing the embedding space. Experiments on public LoRa and Wi-Fi datasets show that HoRFFI achieves absolute improvements of \(0.112\) and \(0.288\) in novel-class accuracy, respectively, and corresponding absolute AUC
improvements of \(0.029\) and \(0.060\) over the best-performing baselines.

\end{abstract}
\begin{IEEEkeywords}
Radio frequency fingerprint identification, High-openness recognition, Variational information bottleneck, IoT device authentication.
\end{IEEEkeywords}

\section{Introduction}

Rapid advances in wireless communication technologies have accelerated the widespread adoption of the Internet of Things (IoT) and the growth of connected devices \cite{zhang2025physical}. This expansion broadens the attack surface and increases the risks of unauthorized access and identity spoofing \cite{aouedi2024survey}. Device authentication has therefore become essential to IoT security \cite{liu2021machine}. Conventional authentication methods generally rely on cryptographic keys or software-based identifiers. The former introduce key-management overhead and remain vulnerable to key leakage, whereas the latter may be tampered with or spoofed \cite{xie2025novel,li2023class}.
In this context, radio frequency fingerprint identification (RFFI) has emerged as a promising physical-layer device authentication technique \cite{zhang2021radio,zhao2025survey}. RFFI derives device identity from distinctive signal characteristics arising from manufacturing variations in radio frequency (RF) transceivers \cite{wang2025review}. Because these characteristics are inherently tied to the transmitter hardware and difficult to replicate precisely, RFFI reduces reliance on transferable digital credentials and makes identity spoofing more difficult \cite{jagannath2022comprehensive,shen2022towards}.

 Recent advances in deep learning have enabled RFFI systems to learn more discriminative device representations, substantially improving identification accuracy\cite{yu2019robust,xie2021generalizable,cai2024toward,cai2025open}.
Despite these advances, most open-set RFFI studies primarily focus on known-device classification and unknown-device rejection \cite{huang2024radio,wu2024open,yin2024multi}. Several scalable RFFI frameworks further support post-training device enrollment without retraining the feature extractor \cite{shen2022towards,xie2025novel}.
However, the impact of limited training-class diversity on the generalization of the learned representation has received little attention. 
We consider a practical deployment setting with only a few labeled base-device classes and a feature extractor that remains fixed after training. During deployment, newly authorized devices must be enrolled and subsequently identified, whereas unregistered devices must be rejected.
Consequently, the number of device classes encountered during deployment may substantially exceed the number of base-device classes available during training, giving rise to what we term a high-openness RFFI setting. Under this setting, representations learned from only a few base-device classes may become overly specialized to those classes, limiting their transferability to device classes unseen during training.

To address this challenge, we propose HoRFFI, a High-Openness Radio Frequency Fingerprint Identification framework comprising training, enrollment, and testing stages. HoRFFI supports post-training device enrollment, enrolled-device identification, and unknown-device rejection without retraining the feature extractor. During training, the extractor is optimized using a Similarity-Enhanced Variational Information Bottleneck (SVIB) objective. SVIB augments the standard VIB objective, which promotes compact, task-relevant representations \cite{alemi2016deep}. It further incorporates similarity pseudo-labels derived from stochastic feature augmentation and clustering, thereby providing pairwise supervision that complements supervision from the few base-device classes. This similarity-aware supervision reduces over-specialization to the few training classes and promotes a more transferable embedding space. After training, the extractor remains fixed throughout enrollment and testing.

HoRFFI differs from existing methods in several aspects. 
Conventional deep learning-based RFFI assumes identical training and testing device classes \cite{shen2021spectrogram,xie2025towards}, whereas many open-set RFFI methods primarily focus on rejecting unseen devices and do not support their subsequent enrollment as authorized identities \cite{chen2023extreme,sun2026enhancing}. 
ScalableRFFI and MLGPN address this limitation by supporting enrollment without retraining the feature extractor. ScalableRFFI employs a triplet-loss-trained extractor, embedding-based enrollment, and a \(k\)-nearest neighbor (\(k\)-NN) rule for identification and rejection \cite{shen2022towards}, whereas MLGPN uses episodic meta-learning and Gaussian prototypes to address few-shot open-set recognition \cite{xie2025novel}. However, neither method explicitly investigates representation learning with only a few base-device classes.
Class-incremental specific emitter identification (SEI) methods instead perform explicit model adaptation as new emitter classes arrive \cite{li2023class,shen2025class}. 
In contrast, HoRFFI performs no class-incremental optimization after deployment. It combines post-training enrollment with an SVIB-based representation objective
designed for generalization from only a few base-device classes.

The main contributions of this paper are summarized as follows:
\begin{itemize}
    \item We formulate a high-openness RFFI problem under limited training-class diversity. In this setting, a feature extractor trained exclusively on a small number of labeled base-device classes must support post-training device enrollment, enrolled-device identification, and unknown-device rejection without retraining.   

    \item We propose HoRFFI, whose core SVIB objective augments limited base-device classes supervision with inter-sample similarity pseudo-labels, thereby promoting a more transferable embedding space for post-training class expansion and open-set identification.

    \item Comprehensive experiments on LoRa and Wi-Fi datasets under different high-openness settings demonstrate that HoRFFI consistently improves newly enrolled-device identification while maintaining competitive or superior unregistered-device rejection performance compared with representative methods.
    
\end{itemize}

The remainder of this paper is structured as follows. Section~\ref{sec:relatedW} reviews the related work. Section~\ref{sec:Methodly} formulates the high-openness RFFI problem, introduces the system model, and presents the design of HoRFFI and its SVIB objective. Section~\ref{sec:Experiments} reports the experimental evaluation. Finally, Section~\ref{sec:Conclusion} concludes the paper.

\section{Related Work}
\label{sec:relatedW}
This section reviews the literature most relevant to HoRFFI from three perspectives: deep learning-based RFFI, open-set and open-world recognition, and information bottleneck-based representation learning.
 \subsection{Deep Learning-Based RFFI} 
The rapid advancement of deep learning has substantially improved feature learning and identification performance in RFFI. A variety of deep neural network architectures have been applied to RFFI, including  convolutional neural networks (CNNs) \cite{shen2022towards,shen2026towards}, long short-term
memory networks (LSTMs) \cite{peng2024channel}, generative adversarial
networks (GANs) \cite{jiang2024radio}, and Transformers \cite{shen2021radio}. However, most existing RFFI methods assume a closed‑set where all classes observed during testing are available during training. This assumption limits their applicability to practical deployments in which newly authorized devices may be introduced after training.

 \subsection{Open-Set and Open-World Recognition}
 Open-set recognition (OSR) \cite{geng2020recent} requires a model to identify samples from known classes while rejecting samples from classes unseen during training. In RFFI, Xie et al. \cite{xie2021generalizable} improve feature separability through hyperspherical projection and cosine-distance-based matching, while Huang et al. \cite{huang2024radio} combine the Fourier synchrosqueezing transform with supervised contrastive learning to improve open-set authentication under noisy channels. These methods mainly focus on known-device identification and unknown-device rejection.
 
 Several scalable RFFI frameworks further support post-training device enrollment. ScalableRFFI \cite{shen2022towards} combines a triplet-loss-trained feature extractor with $k$-NN-based enrollment and
identification, whereas MLGPN \cite{xie2025novel} employs episodic meta-learning and Gaussian prototypes for few-shot open-set recognition. 
 Open-world recognition (OWR) \cite{bendale2015towards} goes beyond OSR by discovering novel classes and incrementally updating the recognition model. Related methods include ORCA \cite{cao2021open}, which reduces bias toward known classes through an uncertainty-adaptive margin, and OpenNCD \cite{liu2023openworld}, which uses bi-level contrastive learning for novel-class discovery. In RFFI, OpenRFI \cite{han2025open} proposes the requirement to classify unknown categories as well, but in specific implementation, this scheme assumes that the total number of known and unknown categories is known. 
 However, these studies do not explicitly investigate the transferability of learned representations under the high-openness setting considered in this work, where only a few base-device classes are available for training.


\subsection{Information Bottleneck-Based Representation Learning} 
The Information Bottleneck (IB) \cite{tishby2000information} principle provides a theoretical framework for learning representations that balance compression and prediction capabilities. The deep variational information bottleneck (VIB) \cite{alemi2016deep} makes IB applicable to deep neural networks and has been widely used to improve representation compactness and generalization. Cen et al. \cite{cen2023enlarging} further investigated open-set recognition from an IB perspective by preserving instance-specific and class-specific information. However, these works are primarily focused on image recognition and have not been explored in the RFFI domain. Moreover, they do not address the high-openness scenario where newly authorized device classes must be enrolled and subsequently identified, while unregistered devices remain rejected. Nevertheless, IB provides a principled paradigm for learning transferable representations by controlling the information retained in the embedding space. Our SVIB objective extends VIB with similarity-aware supervision to improve representation transferability under limited base-device classes diversity.

\section{Methodology}
\label{sec:Methodly}

\subsection{Problem Formulation}
We consider a high-openness RFFI scenario where the training set \(\mathcal{D}_{\text{train}} = \{(\mathbf{x}_i, y_i)\}_{i=1}^{N}\) contains samples from only \(N_{\text{CT}}\) base-device classes, with \(y_i \in \{1, \ldots, N_{\text{CT}}\}\). During deployment, the system encounters a test set \(\mathcal{D}_{\text{test}}\) containing samples from \(N_{\text{CE}}\) total classes, where \(N_{\text{CE}} \gg N_{\text{CT}}\). The additional \(N_{\text{CE}} - N_{\text{CT}}\) classes are unseen during training and after training, a subset of the unseen classes may be enrolled as authorized novel classes and must be identified, whereas classes that remain unregistered must be rejected. The openness of this setting is quantified as \cite{scheirer2012toward}:
\begin{equation}
\text{Openness} = 1 - \sqrt{\frac{2 \times N_{\text{CT}}}{N_{\text{CE}} + N_{\text{CR}}}}, 
\label{eq:openness}
\end{equation}
where \(N_{\text{CT}}\) is the number of classes used in training, \(N_{\text{CE}}\) is the number of classes used in evaluation (testing), and \(N_{\text{CR}}\) is the number of classes to be recognized. 

Our objective is to learn an encoder \(f_\theta: \mathcal{X} \to \mathcal{Z}\) that maps inputs to a discriminative embedding space. Following the scalable enrollment and identification procedure in \cite{shen2022towards}, a distance-based \(k\)-NN rule is applied to the frozen embeddings to identify enrolled devices and reject unregistered ones. The encoder must support these tasks even when only \(N_{\text{CT}}\) base-device classes are available during training.

\subsection{Overview of the HoRFFI Framework}

\begin{figure}[t]
    \centering
    \includegraphics[width=0.48\textwidth]{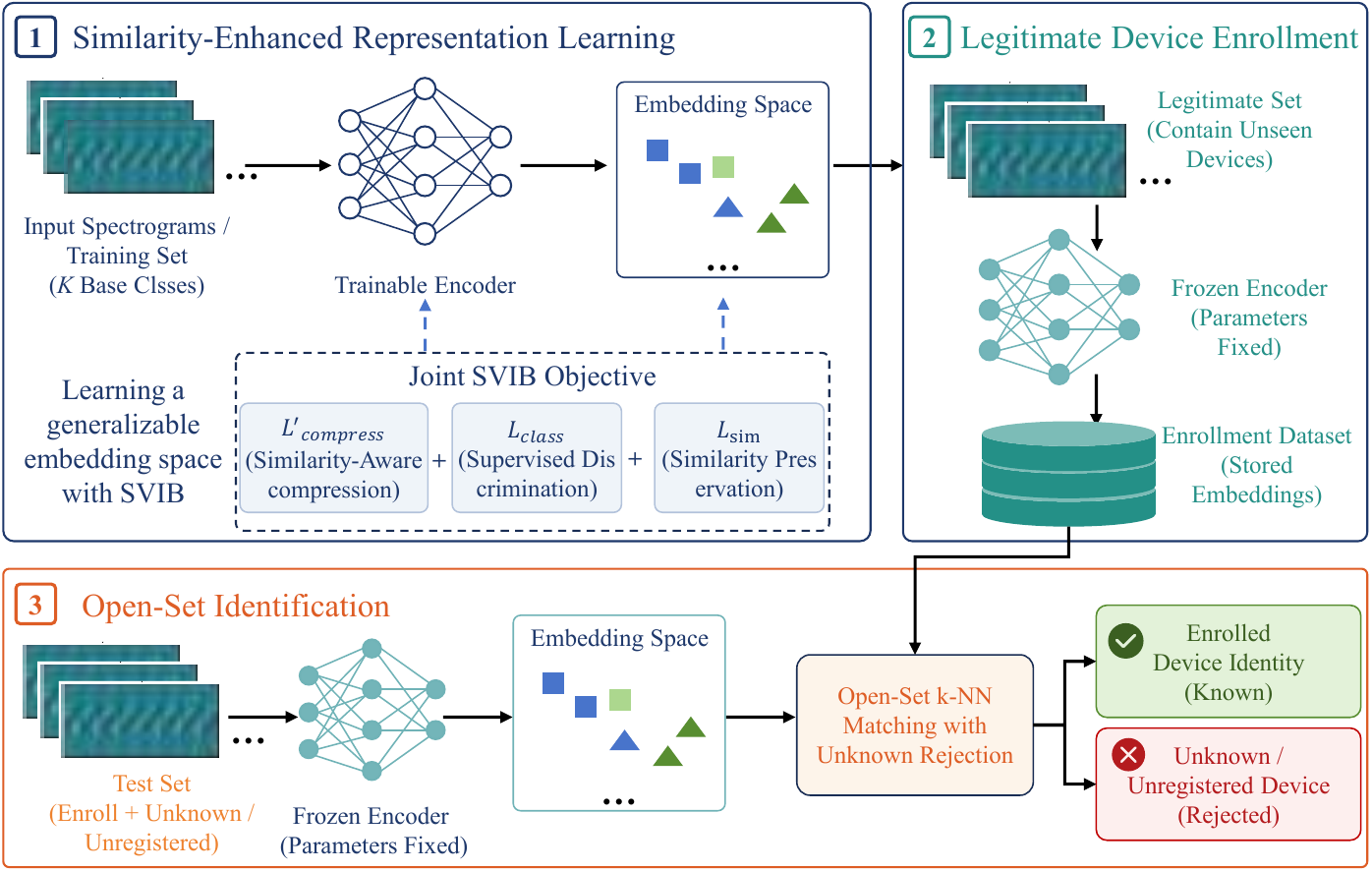}
    \caption{
    Overview of the HoRFFI framework. The encoder is trained with the SVIB objective (Stage 1), then frozen for subsequent enrollment (Stage 2) and $k$-NN-based open-set identification (Stage 3). The framework supports device enrollment without retraining.
    }
    \label{fig:HORFFI_overview}
\end{figure}

The overall architecture of the proposed HoRFFI framework is illustrated in Fig.~\ref{fig:HORFFI_overview}, which operates in three stages: similarity-enhanced representation learning (training), legitimate device enrollment (enrollment), and open-set identification (testing). In the training stage, raw I/Q samples are transformed into spectrograms and fed into the encoder, which is optimized via the proposed SVIB objective. The encoder learns a representation space that is both discriminative for base classes and generalizable to the novel device classes. In the enrollment stage, newly authorized devices are registered by extracting and storing their embeddings using the frozen encoder, enabling dynamic device addition without retraining. In the testing stage, each incoming sample is mapped to a query embedding and compared against all enrollment embeddings via $k$-NN classification to determine its identity.

\subsection{Similarity-Enhanced Variational Information Bottleneck}

Standard VIB acts as a global filter, compressing all information irrelevant to base-class classification. In open-set RFFI, however, this is detrimental because novel classes require similarity structures that are irrelevant to base-class labels. Our key insight is to compress information conditioned on similarity structures $Y_s$, rather than unconditionally.

Ideally, the Markov chain should follow \((Y_t, Y_s) \rightarrow X \rightarrow Z\), where \(Y_t\) denotes ground-truth labels and \(Y_s\) captures similarity information. Since \(Y_s\) is not directly observable, we introduce an additional stochastic node \(\hat{Z}\) obtained via feature mixup, yielding the Markov structure:
\begin{equation}
Y_t \rightarrow X \rightarrow Z \rightarrow \hat{Z} \rightarrow Y_s, \quad
Y_t \rightarrow X \rightarrow Z \rightarrow Y_p, 
\label{eq:svib_markov}
\end{equation}
where \(Y_p\) is the model's predicted label and \(Y_s\) is estimated from \(\hat{Z}\) via clustering. 

\begin{figure}[t]
    \centering
    \includegraphics[width=0.48\textwidth]{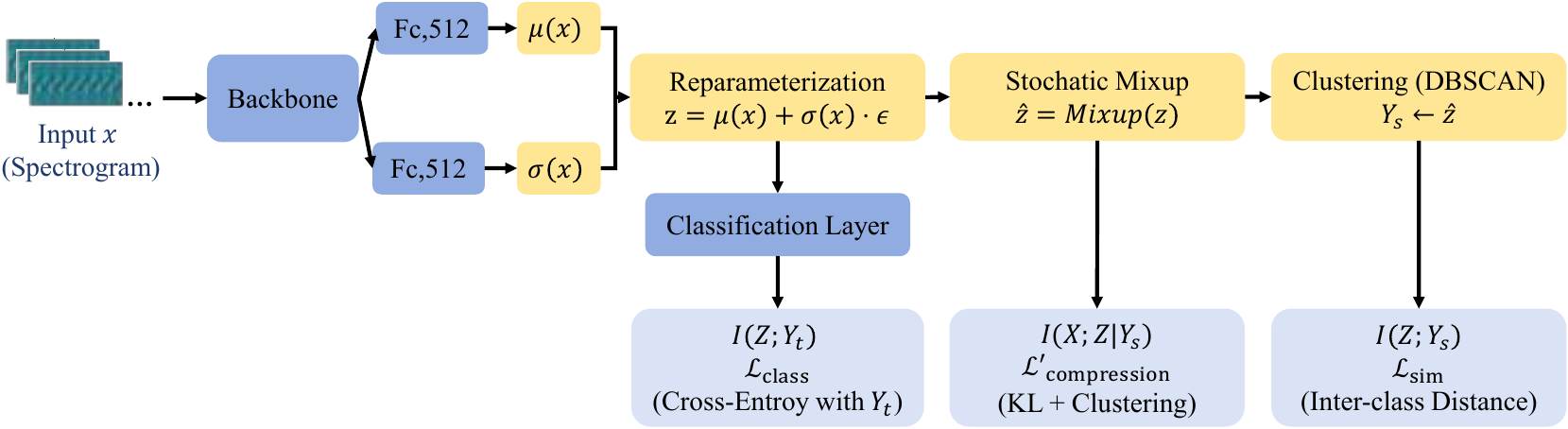}
    \caption{Detailed training pipeline of the SVIB encoder. The encoder has two output branches that separately predict the mean $\mu$ and standard deviation $\sigma$ of the latent distribution. Reparameterization yields \(z\), which undergoes stochastic mixup to produce \(\hat{z}\). Clustering on \(\hat{z}\) generates similarity pseudo-labels \(Y_s\). The three losses jointly optimize the encoder: classification loss \(\mathcal{L}_{\text{class}}\), compression loss \(\mathcal{L}^{\prime}_{\text{compression}}\), and similarity loss \(\mathcal{L}_{\text{sim}}\).}
    \label{fig:svib_encoder_detail}
\end{figure}

Fig. \ref{fig:svib_encoder_detail} shows the detailed training pipeline of the SVIB encoder. The SVIB objective is derived from the information bottleneck principle and implemented through three surrogate objectives, including similarity-aware compression, supervised discrimination, and similarity preservation:
\begin{equation}
\min \mathcal{L} = \beta_1 I(X;Z|Y_s) - \beta_2 I(Z;Y_t) - \beta_3 I(Z;Y_s).
\label{eq:svib_objective}
\end{equation}
This formulation motivates that only information irrelevant to similarity is compressed, while both discriminative and similarity-aware information are preserved.

\subsubsection{Similarity-Aware Compression}

The compression term in SVIB differs fundamentally from standard VIB: rather than compressing all mutual information between \(X\) and \(Z\), we compress only the information unrelated to similarity. Under the similarity variable \(Y_s\), this conditional compression is expressed as:
\begin{equation}
I(Z;X|Y_s)
=
\mathbb{E}_{x\sim p(x|Y_s)}
\Big[
D_{\mathrm{KL}}\big(
p(z|x, Y_s)
\|p(z|Y_s)
\big)
\Big].
\end{equation}
Since the marginal \(p(z|Y_s)\) is intractable, we introduce a variational approximation \(r(z|Y_s)\), yielding the upper bound:
\begin{equation}  
I(Z;X|Y_s)
\le
\mathbb{E}_{x\sim p(x|Y_s)}
\Big[
D_{\mathrm{KL}}\big(
p(z|x, Y_s)
\|r(z|Y_s)
\big)
\Big].
\label{eq:KLandIZXYs}
\end{equation}

A key distinction from standard VIB lies in the choice of prior. In VIB, \(r(z) = \mathcal{N}(0, I)\) works with softmax classifiers because they treat samples independently. In open-set scenarios, however, distance-based classifiers like $k$-NN rely on relative sample relationships — pulling all samples to the origin would distort these relationships. We therefore adopt a data-driven prior \(r(z|Y_s) = \mathcal{N}(\mu, I)\), where \(\mu\) encodes similarity information.

Since \(Y_s\) is unobserved, the ideal encoder \(p(z|x, Y_s)\) is not directly available. We introduce a parametric encoder \(p_{\phi, h}(z|x)\) that depends only on \(x\), where \(\phi\) denotes the network parameters and \(h\) represents a similarity estimation mechanism (instantiated as clustering on \(\hat{Z}\)).  The resulting cluster assignments provide an auxiliary pseudo-label $Y_s$, which is not a ground-truth semantic variable but serves as a training signal to guide similarity-aware regularization. This yields the practical compression loss:
\begin{equation}
\begin{split}
&\mathcal{L}^{\prime}_{\mathrm{compression}}
=\\&~~~~~~~~~~~~~
\mathbb{E}_{y_s\sim p(y_s)}
\mathbb{E}_{x\sim p(x\mid y_s)}
\left[
D_{\mathrm{KL}}\!\left(
p_{\phi,h}(z\mid x)
\Vert
r(z\mid y_s)
\right)
\right].
\label{eq:compression_loss_prime}
\end{split}
\end{equation}

To make this tractable, we adopt the reparameterization trick, where \(z = \mu(x) + \sigma(x)\epsilon\) and \(\epsilon \sim \mathcal{N}(0, I)\) \cite{kingma2014autoencoding}. This necessitates that the encoder outputs both the mean \(\mu(x)\) and variance \(\sigma^2(x)\). In practice, we obtain pseudo-labels via clustering with DBSCAN on \(\hat{Z}\) and set \(r(z|Y_s=c)=\mathcal{N}(\mu_c, I)\), where \(\mu_c\) is the cluster center. The KL divergence then decomposes as:
\begin{equation}
D_{\mathrm{KL}}
=
\frac{1}{2}
\sum_{k}
\left(
\sigma_k^2(x)
-
\log \sigma_k^2(x)
-
1
\right)
+
\frac{1}{2}
\|\mu(x)-\mu_c\|^2.
\label{eq:kl_decomp}
\end{equation}

The decomposition in Eq.~\eqref{eq:kl_decomp} reveals an important insight: the KL divergence enforces both mean alignment and variance regularization. However, in our implementation, we decouple these two effects. The variance term is directly regularized by the KL divergence, while the mean alignment is achieved through clustering. For samples \(\{z_i\}_{i=1}^{N_c}\) in cluster \(c\) with center \(\mu_c\), the total intra-cluster pairwise distance satisfies:
\begin{equation}
\sum_{i<j} \|z_i - z_j\|^2 = N_c \sum_{i=1}^{N_c} \|z_i - \mu_c\|^2.
\label{eq:cluster_identity}
\end{equation}
Thus, minimizing pairwise distances is equivalent to minimizing the distance to the center:
\begin{equation}
\mathbb{E}\|z_i - \mu_c\|^2 = \|\mu(x_i) - \mu_c\|^2 + \mathrm{Tr}\bigl(\mathrm{diag}(\sigma^2(x_i))\bigr).
\end{equation}
Since variance is already regularized by the KL term, minimizing intra-cluster distance effectively minimizes \(\|\mu(x_i) - \mu_c\|^2\), achieving the same mean-alignment objective. The final surrogate loss for compression is:
\begin{equation}
\label{eq:L1_compress}
\begin{split}
\mathcal{L}^{\prime}_{\text{compression}}
=
k_1 & \sum_{k=1}^{K} \bigl( \sigma_k^2(x) - \log \sigma_k^2(x) - 1 \bigr) \\
&+ k_2 \sum_{i<j, \; Y_s^{z_i}=Y_s^{z_j}} \|z_i - z_j\|^2 .
\end{split}
\end{equation}

\subsubsection{Supervised Discrimination}

While compression ensures intra-class compactness, it alone cannot guarantee that different classes are separable. To preserve discriminative information, we maximize the mutual information between \(Z\) and the ground-truth label \(Y_t\). Following the variational information bottleneck framework, we introduce a classifier \(q_\psi(y_t|z)\) and optimize the lower bound:
\begin{equation}
I(Z;Y_t) \ge \mathbb{E}_{p(x, y_t)} \mathbb{E}_{p_{\phi, h}(z|x)} \left[ \log q_\psi(y_t|z) \right].
\end{equation}
This corresponds to minimizing the standard cross-entropy loss:
\begin{equation}
\mathcal{L}_{\text{class}} = \mathbb{E}_{(x, y_t)} \left[ - \log q_\psi(y_t|z) \right].
\end{equation}
Although this term is identical to the supervised loss in standard VIB, it serves a critical role in our framework: it ensures the latent space retains class-discriminative information, which is essential for the $k$-NN classifier in the open-set identification stage. Together with \(\mathcal{L}^{\prime}_{\text{compression}}\), it encourages a space that is both compact within each class and well-separated across classes.

\subsubsection{Similarity Preservation}

\textbf{Stochastic Mixup.}
A central challenge in similarity-aware representation learning is that the similarity variable \(Y_s\) is not directly observable and must be derived from the learned representations. If \(Y_s\) is a deterministic function of \(Z\), then \(H(Y_s\mid Z)=0\), and the mutual information reduces to
\begin{equation}
I(Z;Y_s)=H(Y_s).
\end{equation}
In this case, the similarity label is completely determined by the current feature representation, and the supervised signal no longer provides additional statistical constraints. The optimization process can only reinforce the existing feature representation structure.

To introduce non-deterministic similarity supervision, we apply stochastic feature-level Mixup and construct the Markov chain
\(Z\rightarrow\hat{Z}\rightarrow Y_s\). Specifically, for each mini-batch representation \(Z_i\), we sample \(\lambda\sim\mathrm{Beta}(\alpha,\alpha)\) and \(M\sim\mathrm{Bernoulli}(p)\). Given a random permutation \(\pi\), the perturbed representation is constructed as
\begin{equation}
\hat{Z}_i =
M\left[\lambda Z_i+(1-\lambda)Z_{\pi(i)}\right]
+(1-M)Z_i.
\label{eq:zhat_mixup}
\end{equation}

Let \(Z_{c,i}\) and \(Z_{c,j}\) denote two representations sharing the same ground-truth class \(c\). Although they tend to receive the same
cluster assignment, stochastic Mixup transforms them into random representations \(\hat{Z}_{c,i}\) and \(\hat{Z}_{c,j}\). When the induced perturbations cross clustering boundaries with nonzero probability, different realizations may yield different cluster assignments. The cluster-derived similarity variable \(Y_s\) is therefore no longer a deterministic function of the original representations, resulting in \(H(Y_s\mid Z)>0\). This conditional uncertainty enables \(Y_s\) to provide an additional similarity-aware signal for regularizing the feature space.

In practice, this behavior is jointly controlled by the Mixup strength, the mixing probability, the geometry of the paired
representations, and the clustering granularity. Under this condition, the resulting similarity assignments provide non-degenerate supervision for regularizing the feature space.

\textbf{Promoting Mutual Information via Inter-Class Separation.}
Let \(D_{ij}=\|z_i-z_j\|_2\)  denote the Euclidean distance between features.
Following \cite{wang2011information},
the differential entropy of \(Z\) can be estimated from the average
logarithm of pairwise distances as
\begin{equation}
\widehat{H}(Z)
=
\frac{d}{n(n-1)}
\sum_{i=1}^{n}
\sum_{\substack{j=1\\j\neq i}}^{n}
\log D_{ij}^{2}
+C
\overset{\mathrm{c}}{=}
\sum_{i=1}^{n}
\sum_{\substack{j=1\\j\neq i}}^{n}
\log D_{ij},
\label{eq:hatHZ}
\end{equation}
where \(n\) is the number of feature representations, \(d\) is the
embedding dimension, and \(C\) is independent of the learned
representations. The notation \(\overset{\mathrm{c}}{=}\) denotes
equality up to a positive multiplicative and/or additive constant.
We assume \(D_{ij}>0\) for \(i\neq j\).

Decomposing the total pairwise log-distance into intra-class and inter-class components with respect to similarity labels \(Y_s\):
\begin{equation}
\begin{aligned}
\sum_{i\neq j}\log D_{ij}
={}&
\sum_{\substack{i\neq j\\Y_s^{z_i}=Y_s^{z_j}}}\log D_{ij}
+
\sum_{\substack{i\neq j\\Y_s^{z_i}\neq Y_s^{z_j}}}\log D_{ij}.
\end{aligned}
\label{eq:distance_decomposition}
\end{equation}

Since \(I(Z;Y_s) = H(Z) - H(Z|Y_s)\), subtracting the intra-class contribution isolates the inter-class separation term. 
Thus, \(I(Z;Y_s)\) and inter-class log-distance follow the same
optimization direction.
Since \(D_{ij}\) and \(\log D_{ij}\) are monotonically related for \(D_{ij}>0\), we use \(D_{ij}\) as a tractable surrogate for the log-distance term.

A critical design choice is using \(Y_s\) rather than \(Y_t\) for inter-class separation. In open-set scenarios, relying solely on training labels \(Y_t\) causes the model to overfit to base classes, limiting generalization to novel devices. Using similarity pseudo-labels \(Y_s\) encourages the model to capture intrinsic similarity structures that transfer across classes. The corresponding surrogate loss is:
\begin{equation}
\mathcal{L}_{\text{sim}} = k_3 \cdot \max\left[
\sum_{Y_s^{z_i} \neq Y_s^{z_j}} m - \|{z}_i - {z}_j\| , \; 0
\right]^2, 
\label{eq:L3_sim}
\end{equation}
where \(m\) is a margin hyperparameter preventing unbounded feature growth.

Overall the loss function of our SVIB is a combination of similarity-aware compression, supervised discrimination, and similarity preservation as:
\begin{equation}
\mathcal{L}_{\mathrm{SVIB}}
=
\beta_1\mathcal{L}^{\prime}_{\mathrm{compression}}
+
\beta_2\mathcal{L}_{\mathrm{class}}
+
\beta_3\mathcal{L}_{\mathrm{sim}},
\label{eq:overall_svib_loss}
\end{equation}
where \(\beta_1\), \(\beta_2\), and \(\beta_3\)  control the contributions of the three objectives, respectively. By jointly optimizing these terms, SVIB learns a representation space that preserves class-discriminative information and sample-level similarity structures while suppressing redundant information.

\section{Experiments}
\label{sec:Experiments}

\subsection{Experimental Setup}

This subsection presents the datasets and preprocessing pipeline, evaluation protocols and metrics, compared methods, and implementation details used in our experiments.

\subsubsection{Datasets and Preprocessing}

We evaluate the proposed method on two public RF fingerprinting datasets, namely the LoRa dataset~\cite{shen2022towards} and the Wi-Fi dataset~\cite{sankhe2019oracle}. For the LoRa dataset, we use a subset of 30 device classes. The Wi-Fi dataset contains signals transmitted by 16 devices.
Unless otherwise specified, seven devices are used as the base-device classes for feature-extractor training (\(N_{\mathrm{CT}}=7\)), with 800 training samples selected from each class.
The remaining devices are excluded from feature-extractor training and are used to evaluate post-training identification.

The preprocessing pipeline follows \cite{shen2022towards} and is applied consistently to both datasets. First, additive white Gaussian noise (AWGN) is introduced into the training samples for data augmentation. Each signal is then normalized according to its root-mean-square (RMS) power. Subsequently, the normalized I/Q samples are transformed into time-frequency spectrograms using the short-time Fourier transform (STFT), followed by the construction of Channel Independent Spectrograms (CISs). Finally, spectral cropping is applied to retain the central \(40\%\) of the frequency band. All compared methods use the same preprocessing pipeline and identical input data to ensure a fair comparison.





\subsubsection{Evaluation Protocols and Metrics}

Following the problem formulation, we distinguish three roles for device classes. The base-device class set \(\mathcal{C}_{\mathrm{base}}\), with \(|\mathcal{C}_{\mathrm{base}}|=N_{\mathrm{CT}}\), is used to train the feature extractor. Classes absent during feature-extractor training may subsequently be enrolled as authorized novel classes, forming the set
\(\mathcal{C}_{\mathrm{novel}}\), or remain unregistered, forming the set \(\mathcal{C}_{\mathrm{unreg}}\). The recognized class set is therefore \(
\mathcal{C}_{\mathrm{rec}}
=
\mathcal{C}_{\mathrm{base}}
\cup
\mathcal{C}_{\mathrm{novel}},
\) with \(
N_{\mathrm{CR}}
=
|\mathcal{C}_{\mathrm{rec}}|\) and \(N_{\mathrm{CE}}
=
|\mathcal{C}_{\mathrm{rec}}
\cup
\mathcal{C}_{\mathrm{unreg}}|.
\)
After training, the feature extractor remains frozen throughout enrollment and testing. Unless otherwise specified, \(500\) samples from each authorized device are used for enrollment, and the enrollment and testing samples are strictly disjoint.

We define two evaluation protocols, as summarized in Table~\ref{tab:evaluation_protocols}. Protocol~I evaluates open-set identification with both enrolled and unregistered devices. For LoRa,
devices \(0\)--\(6\) form \(\mathcal{C}_{\mathrm{base}}\),
devices \(7\)--\(19\) form
\(\mathcal{C}_{\mathrm{novel}}\), and
devices \(20\)--\(29\) form
\(\mathcal{C}_{\mathrm{unreg}}\). Accordingly,
\(N_{\mathrm{CT}}=7\),
\(N_{\mathrm{CR}}=20\), and
\(N_{\mathrm{CE}}=30\).
For Wi-Fi, devices \(0\)--\(6\),
\(7\)--\(10\), and
\(11\)--\(15\) form the base, enrolled novel, and unregistered class sets, respectively, yielding
\(N_{\mathrm{CT}}=7\),
\(N_{\mathrm{CR}}=11\), and
\(N_{\mathrm{CE}}=16\).

Protocol~II evaluates post-training class expansion under different high-openness conditions. For LoRa, the feature extractor is trained using devices \(0\)--\(4\), \(0\)--\(5\), or \(0\)--\(6\), corresponding to \(N_{\mathrm{CT}}\in\{5,6,7\}\). All \(30\) device classes are then enrolled and evaluated, such that \(N_{\mathrm{CR}}=N_{\mathrm{CE}}=30\).
For Wi-Fi, the same training configurations are used, while all \(16\) classes are enrolled and evaluated, yielding \(N_{\mathrm{CR}}=N_{\mathrm{CE}}=16\).
Thus, Protocol~II varies the discrepancy between the number of classes used for representation learning and the number of classes subsequently enrolled and identified, without including unregistered test classes.

\begin{table}[t]
\centering
\footnotesize
\caption{Device-class partitions under the evaluation protocols.}
\label{tab:evaluation_protocols}
\resizebox{\columnwidth}{!}{
\begin{tabular}{c|c|c|c|c}
\toprule
Protocol & Dataset & Base Classes & Enrolled Classes & Unregistered Classes \\
\midrule
I  & LoRa  & 0--6 & 0--19 & 20--29 \\
I  & Wi-Fi & 0--6 & 0--10 & 11--15 \\
II & LoRa  & 0--4 / 0--5 / 0--6 & 0--29 & None \\
II & Wi-Fi & 0--4 / 0--5 / 0--6 & 0--15 & None \\
\bottomrule
\end{tabular}}
\end{table}


Across both protocols, we report enrolled-device accuracy, denoted by
\(\mathrm{ACC}\), to measure identification performance over all
registered devices. Under Protocol~I, we additionally report
\(\mathrm{ACC}_{\mathrm{base}}\) and
\(\mathrm{ACC}_{\mathrm{novel}}\), which are computed over the base
classes used for feature-extractor training and the novel classes
subsequently enrolled after training, respectively. We further report
the novelty gap, defined as
$
\mathrm{Novelty\ Gap}
=
\mathrm{ACC}_{\mathrm{base}}
-
\mathrm{ACC}_{\mathrm{novel}}
,
$
to quantify the performance disparity between the two groups.
Unregistered-device rejection is evaluated using the area under the receiver operating characteristic curve (AUC) and the equal error rate (EER). Under Protocol~II, all testing devices have been enrolled, and \(\mathrm{ACC}\) is therefore used as the primary metric for evaluating identification performance under different openness levels.

All results are reported as the mean and standard deviation over five independent runs with different random seeds ($42$ to $46$).
\subsubsection{Compared Methods}
\label{sec:compared_methods}

We conduct comparisons at both the framework and loss-function levels.
At the framework level, HoRFFI is compared with ScalableRFFI~\cite{shen2022towards} and MLGPN~\cite{xie2025novel}. ScalableRFFI employs triplet loss and a
\(k\)-NN classifier to support post-training device enrollment,
identification, and unregistered-device rejection. MLGPN employs
episodic meta-learning and an open-set loss to learn Gaussian
prototype-based representations, with identification and rejection
performed using the Mahalanobis distance.

At the loss-function level, SVIB is compared with five representative objectives. Contrastive loss~\cite{hadsell2006dimensionality} minimizes distances between same-class samples while separating different-class samples by a predefined margin. SupCon~\cite{khosla2020supervised} contrasts same-class samples against samples from other classes within each training batch. Following~\cite{sun2026knowledge}, N-pair+Center combines multiclass N-pair loss~\cite{sohn2016improved} for inter-class separation with center loss~\cite{wen2016discriminative} for intra-class compactness. Triplet loss~\cite{schroff2015facenet} enforces a margin between anchor-positive and anchor-negative distances and corresponds to the
ScalableRFFI configuration in the framework-level comparison.
VIB~\cite{alemi2016deep} combines label-supervised classification with variational compression to learn compact, task-relevant
representations.

\subsubsection{Implementation Details}
\label{sec:implementation_details}

For the framework-level comparison, all methods use identical preprocessed inputs and device-class partitions. ScalableRFFI and HoRFFI use the same CNN encoder adopted from \cite{shen2022towards}, whereas MLGPN retains its specialized representation and decision mechanisms. The training objective, enrollment representation, and open-set decision rule of each framework are otherwise preserved.

For the loss-function-level comparison, all loss functions are evaluated using the same pipeline as \cite{shen2022towards}. The CNN encoder, training data, enrollment embeddings, \(k\)-NN classifier, and open-set decision rule are kept unchanged. Only the loss function used to train the feature extractor is replaced.

For HoRFFI, the feature extractor is trained using the Adam optimizer with an initial learning rate of 0.001, a batch size of 32, and a maximum of 1000 training epochs. To prevent overfitting and improve convergence, we employ an early stopping strategy with a patience of 20 monitoring the validation loss, and a learning rate reduction scheduler (ReduceLROnPlateau) that decays the learning rate by a factor of 0.2 if the validation loss plateaus for 10 consecutive epochs. The dimension of the learned embedding is set to 512. After training, the feature extractor is frozen, and 500 samples from each authorized device are used for enrollment. During testing, device identification is performed using a \(k\)-NN classifier with \(k=15\), while unregistered-device rejection follows the distance-based decision rule in \cite{shen2022towards}.

All experiments are conducted on a single NVIDIA GeForce RTX 3080 GPU.

\subsection{Comprehensive Evaluation}
\label{sec:comprehensive_evaluation}

\subsubsection{Framework-Level Comparison}
\label{framework level}
We first compare HoRFFI with ScalableRFFI and MLGPN under Protocol~I. Table~\ref{tab:framework_comparison} reports the framework-level comparison on the LoRa and Wi-Fi datasets.

\begin{table*}[t]
\centering
\footnotesize
\caption{Framework-level comparison under Protocol~I on the LoRa and
Wi-Fi datasets. Results are reported as mean $\pm$ standard deviation.
The best and second-best results for each dataset are highlighted in
bold and underlined, respectively.}
\begin{tabular}{clcccccc}
\toprule
Dataset
& Method
& ACC ($\uparrow$)
& $\mathrm{ACC}_{\mathrm{base}}$ ($\uparrow$)
& $\mathrm{ACC}_{\mathrm{novel}}$ ($\uparrow$)
& Novelty Gap ($\downarrow$)
& AUC ($\uparrow$)
& EER ($\downarrow$) \\
\midrule

\multirow{3}{*}{LoRa}
& MLGPN
& $0.662 \pm 0.002$
& $\mathbf{0.962 \pm 0.008}$
& $0.501 \pm 0.003$
& $0.460 \pm 0.010$
& $0.784 \pm 0.040$
& $0.290 \pm 0.036$ \\

& ScalableRFFI
& $\underline{0.732 \pm 0.138}$
& $0.774 \pm 0.120$
& $\underline{0.709 \pm 0.149}$
& $\underline{0.065 \pm 0.043}$
& $\underline{0.794 \pm 0.062}$
& $\underline{0.280 \pm 0.051}$ \\

& HoRFFI
& $\mathbf{0.840 \pm 0.051}$
& $\underline{0.875 \pm 0.082}$
& $\mathbf{0.821 \pm 0.038}$
& $\mathbf{0.054 \pm 0.055}$
& $\mathbf{0.823 \pm 0.030}$
& $\mathbf{0.257 \pm 0.024}$ \\

\midrule

\multirow{3}{*}{Wi-Fi}
& MLGPN
& $\underline{0.622 \pm 0.005}$
& $\underline{0.949 \pm 0.015}$
& $0.049 \pm 0.019$
& $0.900 \pm 0.032$
& $\underline{0.628 \pm 0.015}$
& $\underline{0.404 \pm 0.020}$ \\

& ScalableRFFI
& $0.529 \pm 0.028$
& $0.653 \pm 0.018$
& $\underline{0.311 \pm 0.050}$
& $\mathbf{0.343 \pm 0.038}$
& $0.594 \pm 0.015$
& $0.438 \pm 0.021$ \\

& HoRFFI
& $\mathbf{0.823 \pm 0.012}$
& $\mathbf{0.951 \pm 0.002}$
& $\mathbf{0.599 \pm 0.035}$
& $\underline{0.352 \pm 0.036}$
& $\mathbf{0.688 \pm 0.041}$
& $\mathbf{0.387 \pm 0.044}$ \\

\bottomrule
\end{tabular}
\label{tab:framework_comparison}
\end{table*}

As shown in Table~\ref{tab:framework_comparison}, HoRFFI achieves more
balanced identification performance. On LoRa, it obtains an ACC of
\(0.840\) and an \(\mathrm{ACC}_{\mathrm{novel}}\) of \(0.821\),
representing absolute improvements of \(0.108\) and \(0.112\) over
the strongest corresponding baselines, respectively.
Although its ACC\(_{\rm base}\) is lower than that of MLGPN, the novelty gap decreases from \(0.460\) to \(0.054\). This result shows that the overall improvement of HoRFFI primarily arises from better generalization to newly enrolled devices rather than stronger fitting to the base-device classes. The improvement is more pronounced on Wi-Fi. HoRFFI maintains an ACC\(_{\rm base}\) of \(0.951\), which is comparable to the \(0.949\) achieved by MLGPN, while increasing ACC\(_{\rm novel}\) from \(0.311\), obtained by the strongest baseline for this metric, to \(0.599\). 
Consequently, HoRFFI achieves an overall ACC of \(0.823\),
representing an absolute improvement of \(0.201\) over the strongest
baseline. HoRFFI therefore narrows the performance gap between base
and enrolled novel devices while improving overall identification
accuracy.

HoRFFI also consistently improves unregistered-device rejection. On LoRa, it increases the AUC from \(0.794\) to \(0.823\) and reduces the EER from \(0.280\) to \(0.257\). On Wi-Fi, it increases the AUC from \(0.628\) to \(0.688\) and reduces the EER from \(0.404\) to \(0.387\). Therefore, the improved identification of enrolled novel devices is not achieved at the expense of distinguishing enrolled devices from unregistered ones.

The performance differences can be related to the design of the compared frameworks. MLGPN achieves strong identification performance on the base-device classes, but its substantial degradation on enrolled novel devices indicates that its episodic learning and Gaussian-prototype-based representation do not generalize effectively when the feature extractor is trained using only a few device classes. ScalableRFFI is closer to our framework. However, its triplet supervision remains dependent on the class relationships available among the limited training devices. HoRFFI retains this scalable enrollment pipeline while introducing SVIB to achieve the strongest overall performance among the compared frameworks.

\subsubsection{Loss-Function-Level Comparison}

We compare SVIB with five representative training objectives under Protocol~I. Table~\ref{tab:loss_comparison} reports the results on the LoRa and Wi-Fi datasets.

\begin{table*}[t]
\centering
\footnotesize
\caption{Loss-function-level comparison under Protocol~I on the LoRa
and Wi-Fi datasets. Results are reported as mean $\pm$ standard
deviation. The best and second-best results for each dataset are
highlighted in bold and underlined, respectively.}
\begin{tabular}{clcccccc}
\toprule
Dataset
& Training Objective
& ACC ($\uparrow$)
& $\mathrm{ACC}_{\mathrm{base}}$ ($\uparrow$)
& $\mathrm{ACC}_{\mathrm{novel}}$ ($\uparrow$)
& Novelty Gap ($\downarrow$)
& AUC ($\uparrow$)
& EER ($\downarrow$) \\
\midrule

\multirow{6}{*}{LoRa}
& Triplet
& $0.732 \pm 0.138$
& $0.774 \pm 0.120$
& $0.709 \pm 0.149$
& $0.065 \pm 0.043$
& $0.794 \pm 0.062$
& $\underline{0.280 \pm 0.051}$ \\

& Contrastive
& $0.524 \pm 0.080$
& $0.545 \pm 0.078$
& $0.513 \pm 0.093$
& $\mathbf{0.033 \pm 0.080}$
& $0.662 \pm 0.065$
& $0.387 \pm 0.059$ \\

& N-pair+Center
& $\underline{0.793 \pm 0.045}$
& $\mathbf{0.876 \pm 0.045}$
& $\underline{0.748 \pm 0.046}$
& $0.129 \pm 0.009$
& $\underline{0.795 \pm 0.030}$
& $\underline{0.280 \pm 0.017}$ \\

& SupCon
& $0.501 \pm 0.029$
& $0.577 \pm 0.006$
& $0.461 \pm 0.046$
& $0.116 \pm 0.046$
& $0.715 \pm 0.009$
& $0.347 \pm 0.007$ \\

& VIB
& $0.580 \pm 0.030$
& $0.649 \pm 0.041$
& $0.543 \pm 0.044$
& $0.106 \pm 0.065$
& $0.656 \pm 0.025$
& $0.392 \pm 0.023$ \\

& SVIB
& $\mathbf{0.840 \pm 0.051}$
& $\underline{0.875 \pm 0.082}$
& $\mathbf{0.821 \pm 0.038}$
& $\underline{0.054 \pm 0.055}$
& $\mathbf{0.823 \pm 0.030}$
& $\mathbf{0.257 \pm 0.024}$ \\

\midrule

\multirow{6}{*}{Wi-Fi}
& Triplet
& $0.529 \pm 0.031$
& $0.653 \pm 0.020$
& $0.310 \pm 0.056$
& $\underline{0.343 \pm 0.042}$
& $0.594 \pm 0.017$
& $0.438 \pm 0.024$ \\

& Contrastive
& $0.511 \pm 0.045$
& $0.654 \pm 0.045$
& $0.262 \pm 0.074$
& $0.392 \pm 0.073$
& $0.598 \pm 0.013$
& $0.450 \pm 0.009$ \\

& N-pair+Center
& $\underline{0.790 \pm 0.014}$
& $\underline{0.939 \pm 0.013}$
& $\underline{0.528 \pm 0.029}$
& $0.411 \pm 0.031$
& $\underline{0.684 \pm 0.032}$
& $\mathbf{0.356 \pm 0.027}$ \\

& SupCon
& $0.307 \pm 0.004$
& $0.353 \pm 0.007$
& $0.227 \pm 0.004$
& $\mathbf{0.126 \pm 0.009}$
& $0.582 \pm 0.004$
& $0.419 \pm 0.006$ \\

& VIB
& $0.616 \pm 0.034$
& $0.926 \pm 0.057$
& $0.073 \pm 0.024$
& $0.853 \pm 0.070$
& $0.436 \pm 0.013$
& $0.535 \pm 0.010$ \\

& SVIB
& $\mathbf{0.823 \pm 0.013}$
& $\mathbf{0.951 \pm 0.003}$
& $\mathbf{0.599 \pm 0.039}$
& $0.352 \pm 0.040$
& $\mathbf{0.688 \pm 0.046}$
& $\underline{0.387 \pm 0.049}$ \\

\bottomrule
\end{tabular}
\label{tab:loss_comparison}
\end{table*}

As shown in Table~\ref{tab:loss_comparison}, SVIB achieves the highest ACC and ACC\(_{\rm novel}\) on both datasets. On LoRa, SVIB obtains an ACC of \(0.840\) and an ACC\(_{\rm novel}\) of \(0.821\), outperforming N-pair+Center, the strongest competing objective for both metrics, by 0.047 and 0.073, respectively. Moreover, its ACC\(_{\rm base}\) of \(0.875\) is nearly identical to the best result of \(0.876\). On Wi-Fi, SVIB improves ACC from \(0.790\) to \(0.823\) and ACC\(_{\rm novel}\) from \(0.528\) to \(0.599\), corresponding to  absolute improvements of 0.033 and 0.071, respectively. It also achieves the highest ACC\(_{\rm base}\) of \(0.951\). Although Contrastive and SupCon obtain the smallest novelty gaps on LoRa and Wi-Fi, respectively, they also produce substantially lower ACC\(_{\rm base}\) and ACC\(_{\rm novel}\). Their small gaps therefore arise from similarly limited performance on both groups rather than strong generalization to newly enrolled devices. Among the objectives that achieve competitive identification accuracy, SVIB provides a more favorable balance. These results show that SVIB improves the identification of devices enrolled after training while preserving strong performance on the base-device classes.

SVIB also demonstrates strong unregistered-device rejection performance. On LoRa, it achieves the highest AUC of \(0.823\) and the lowest EER of \(0.257\), increasing AUC by an absolute \(0.028\) and reducing EER by an absolute \(0.023\) over the strongest competing results. On Wi-Fi, SVIB obtains the highest AUC of \(0.688\), although its advantage over N-pair+Center is limited to \(0.004\). N-pair+Center achieves the lowest EER of \(0.356\), while SVIB obtains the second-best result of \(0.387\). Therefore, SVIB provides the strongest enrolled-device identification performance while remaining competitive in unregistered-device rejection.

Since all objectives are evaluated using the same encoder, training data, enrollment procedure, \(k\)-NN classifier, and open-set decision rule, the observed differences can be primarily attributed to their supervision mechanisms. Among the conventional metric-learning objectives, N-pair+Center provides the strongest overall identification performance by jointly promoting intra-class compactness and inter-class separation. However, SVIB achieves higher ACC and ACC\(_{\rm novel}\) on both datasets while maintaining competitive unregistered-device rejection performance. The substantial performance difference between VIB and SVIB further verifies the effectiveness of incorporating inter-sample similarity supervision into the conventional variational information bottleneck.

\subsection{Openness Sensitivity Analysis}
\label{sec:openness_sensitivity}

We next evaluate the compared methods under different high-openness conditions using Protocol~II. This yields openness values ranging from 0.517 to 0.592 on LoRa and from 0.339 to 0.441 on Wi-Fi, as summarized in Table~\ref{tab:openness_config}.

\begin{table}[t]
\centering
\footnotesize
\setlength{\tabcolsep}{4pt}
\caption{Openness configurations for LoRa and Wi-Fi datasets. In our experiments, the scheme is required to recognize all classes in the test set, hence \(N_{\text{CR}} = N_{\text{CE}}\). Openness is computed according to Eq.~\eqref{eq:openness}.}
\begin{tabular}{c|cc|cc}
\toprule
\textbf{Training} & \multicolumn{2}{c|}{\textbf{LORA}} & \multicolumn{2}{c}{\textbf{Wi-Fi}} \\
\textbf{Classes} & Total Classes & Openness & Total Classes & Openness \\
\midrule
5  & 30 & 0.592 & 16 & 0.441 \\
6  & 30 & 0.553 & 16 & 0.388 \\
7  & 30 & 0.517 & 16 & 0.339 \\
\bottomrule
\end{tabular}
\label{tab:openness_config}
\end{table}

\subsubsection{Framework-Level Comparison}

\begin{figure}[t]
    \centering
    \subfloat[LoRa\label{fig:openness_lora_framework}]{
        \includegraphics[width=0.475\linewidth]{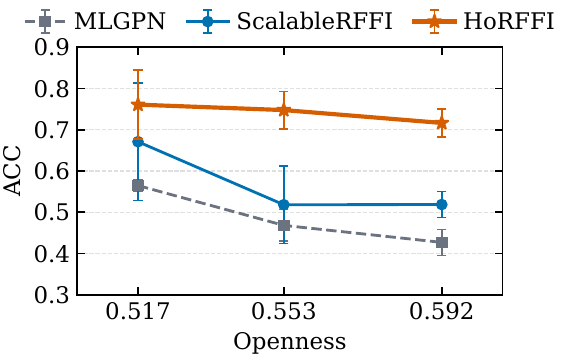}
    }
    \subfloat[Wi-Fi\label{fig:openness_Wi-Fi_framework}]{
        \includegraphics[width=0.475\linewidth]{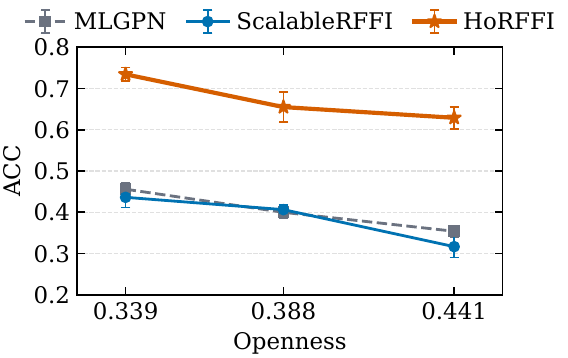}
    }
    \caption{Accuracy of the compared frameworks under different openness levels on the LoRa and Wi-Fi datasets.}
    \label{fig:openness_framework}
\end{figure}

Fig.~\ref{fig:openness_framework} compares the framework-level ACC under different openness levels.

As shown in Fig.~\ref{fig:openness_framework}, HoRFFI consistently achieves the highest ACC across all openness levels on both datasets. On LoRa, as the openness increases from \(0.517\) to \(0.592\), the ACC of HoRFFI decreases moderately from \(0.761\) to \(0.716\), a reduction of only 0.044. 
By comparison, ScalableRFFI decreases from \(0.671\) to \(0.519\), while MLGPN decreases from \(0.565\) to \(0.427\), corresponding to absolute reductions of 0.152 and 0.138, respectively. 
At the highest openness, HoRFFI outperforms ScalableRFFI, the strongest baseline, by 0.198.

The same overall trend is observed on Wi-Fi. HoRFFI achieves ACC values of \(0.734\), \(0.655\), and \(0.629\) at openness levels of \(0.339\), \(0.388\), and \(0.441\), respectively. At the highest openness, the best baseline achieves an ACC of only \(0.354\), whereas HoRFFI maintains an ACC of \(0.629\), yielding an absolute improvement of 0.275. Although all three frameworks exhibit performance degradation as the openness increases, HoRFFI maintains a substantial advantage throughout the evaluated range.

These results reflect the different abilities of the frameworks to support post-training class expansion. MLGPN and ScalableRFFI become less effective when an increasing proportion of the registered devices is absent during feature-extractor training. In contrast, HoRFFI combines scalable device enrollment with SVIB-based representation learning, enabling it to maintain higher overall identification accuracy as the discrepancy between the training and deployment class spaces increases. HoRFFI retains a larger accuracy advantage as the number of base-device classes decreases.

\subsubsection{Loss-Function-Level Comparison}

\begin{figure}[t]
    \centering
    \subfloat[LoRa\label{fig:openness_lora_loss}]{
        \includegraphics[width=0.475\linewidth]{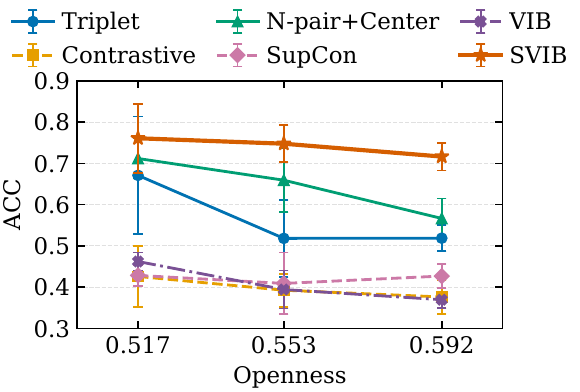}
    }
    \subfloat[Wi-Fi\label{fig:openness_Wi-Fi_loss}]{
        \includegraphics[width=0.475\linewidth]{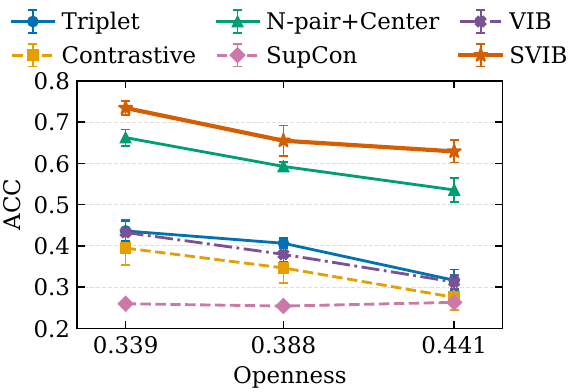}
    }
    \caption{Accuracy of the compared losses under different openness levels on the LoRa and Wi-Fi datasets.}
    \label{fig:openness_loss}
\end{figure}

Fig.~\ref{fig:openness_loss} compares the accuracy achieved by the different training objectives under varying openness levels.

As shown in Fig.~\ref{fig:openness_loss}, SVIB consistently achieves the highest ACC across all openness levels on both datasets. On LoRa, SVIB maintains ACC values of \(0.761\), \(0.748\), and \(0.716\) as the openness increases from \(0.517\) to \(0.592\). In contrast, N-pair+Center, the strongest competing objective, decreases from \(0.712\) to \(0.566\). Consequently, the advantage of SVIB over N-pair+Center increases from 0.049 at the lowest openness to 0.15 at the highest openness.

On Wi-Fi, SVIB achieves ACC values of \(0.734\), \(0.655\), and \(0.629\) at openness levels of \(0.339\), \(0.388\), and \(0.441\), respectively. The corresponding results of N-pair+Center are  \(0.662\), \(0.593\), and \(0.536\). Thus, SVIB achieves consistent absolute ACC improvements of \(0.072\), \(0.062\), and \(0.093\) over the strongest competing objective. Although all competitive objectives experience some performance degradation as the openness increases, SVIB consistently maintains the highest overall identification accuracy.

As openness increases, the performance of all methods generally decreases because the feature extractor is trained using fewer base-device classes relative to the number of devices subsequently enrolled. However, SVIB exhibits a consistently smaller performance degradation and maintains the highest ACC across all openness settings. This result indicates that the similarity-enhanced supervision reduces the feature extractor's sensitivity to training-device diversity, allowing the learned representation space to generalize more effectively when only a small number of device classes are available.

\subsection{Representation Quality Analysis}
\label{sec:representation_quality}

To further examine the learned representation spaces, we evaluate different losses using MAP@R~\cite{musgrave2020metric} and t-SNE under Protocol~II on the LoRa dataset with an openness of \(0.517\). MAP@R is computed over all 30 device classes, while devices 0--3 and 7--14 are selected as representative base and newly enrolled classes, respectively, for visualization. All methods use identical testing samples and t-SNE configurations.

\begin{figure}[t]
    \centering
    \includegraphics[width=0.75\columnwidth]
    {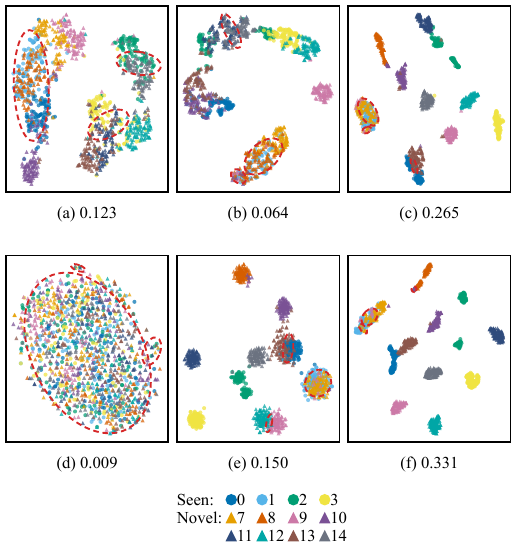}
    \caption{Representation spaces learned using (a) Triplet, (b) Contrastive, (c) N-pair+Center, (d) SupCon, (e) VIB, and (f) SVIB. Values denote overall MAP@R. Circles and triangles represent base and newly enrolled devices, respectively. Red dashed ellipses indicate mixed-label neighborhoods in the t-SNE projection.}
    \label{fig:representation_quality}
\end{figure}

As shown in Fig.~\ref{fig:representation_quality}, SVIB achieves the highest MAP@R of \(0.331\), outperforming N-pair+Center and VIB by \(0.066\) and \(0.181\), respectively. The t-SNE projection provides a qualitative visualization in which SVIB exhibits fewer mixed-label neighborhoods. In contrast, the other losses exhibit varying degrees of class overlap, particularly SupCon. These results demonstrate that incorporating inter-sample similarity supervision improves the neighborhood consistency and retrieval quality of the learned representation space.

\subsection{Ablation Experiment}
\label{sec:ablation}
The experiments in this subsection were conducted using the experimental setup of Protocol~II, and the model was trained with 7 device classes.
\subsubsection{Impact of Loss Components.} Table~\ref{tab:ablation_components} analyzes the contributions of the three loss components in our framework. To investigate their importance, we conduct ablation studies by individually removing each loss or removing any two losses from the SVIB. The configuration \(\mathcal{L}_{\mathrm{class}}+\mathcal{L}_{\mathrm{sim}}\) exhibits substantially greater variability than the full objective, with standard deviations exceeding \(0.3\), highlighting the role of the compression term in stabilizing training. Notably, using either \(\mathcal{L}'_{\mathrm{compression}}\) or \(\mathcal{L}_{\mathrm{sim}}\) without classification supervision produces an ACC close to random guessing, indicating that these components alone cannot establish device-discriminative representations. The complete objective achieves the best performance. This result indicates that, although \(\mathcal{L}_{\mathrm{sim}}\) does not have the ability to independently establish classification, it can further improve the generalization ability of the representation structure and novel classes by utilizing the similarity between samples after compression and category supervision have formed a stable feature space. Therefore, the three components play complementary roles and jointly contribute to the effectiveness of SVIB.

\begin{table}[t]
\centering
\footnotesize
\caption{Ablation results on the LoRa dataset at an openness of \(0.517\). \(\mathcal{L}_{c}\), \(\mathcal{L}_{cls}\), and \(\mathcal{L}_{s}\) denote compression loss, classification loss, and similarity loss, respectively. Best results are highlighted in bold. Results are reported as mean \(\pm\) standard deviation.}
\label{tab:ablation_components}
\begin{tabular}{lcc}
\toprule
Objective
& ACC($\uparrow$)
& ACC$_{\mathrm{novel}}$($\uparrow$)
 \\
\midrule

$\mathcal{L}_{c}$ & $0.0345\pm0.0018$ & $0.0359\pm0.0025$  \\

\(\mathcal{L}_{cls}\)
& \(0.675 \pm 0.238\)
& \(0.664 \pm 0.246\)
 \\

$\mathcal{L}_{s}$ & $0.0349\pm0.0032$ & $0.0357\pm0.0040$  \\

\(\mathcal{L}_{c}+\mathcal{L}_{cls}\)
& \(0.730 \pm 0.139\)
& \(0.718 \pm 0.133\)
 \\

$\mathcal{L}_{c}+\mathcal{L}_{s}$ & $0.0345\pm0.0018$ & $0.0359\pm0.0025$  \\

\(\mathcal{L}_{cls}+\mathcal{L}_{s}\)
& \(0.542 \pm 0.334\)
& \(0.522 \pm 0.338\)
 \\

\(\mathcal{L}_{c}+\mathcal{L}_{cls}+\mathcal{L}_{s}\)
& \(\mathbf{0.761 \pm 0.075}\)
& \(\mathbf{0.740 \pm 0.080}\)
 \\
\bottomrule
\end{tabular}
\end{table}

\subsubsection{Impact of the Number of Enrollment Samples.}

Table~\ref{tab:enrollment_size} examines the trade-off between identification accuracy and recognition latency. Increasing the number of enrollment samples from 100 to 500 improves ACC from \(0.483\) to \(0.761\), while increasing the average \(k\)-NN recognition latency from only \(0.0570\) to \(0.1015\) ms per test sample. This trade-off is particularly suitable for our deployment scenario, where the primary constraint is the limited diversity of physical devices rather than the number of signals obtainable from an available device. We therefore use 500 samples for enrollment throughout the experiments. Applications with stricter latency or enrollment-storage requirements may instead use fewer samples at the cost of reduced identification accuracy.

\begin{table}[t]
\centering
\footnotesize
\caption{Impact of the number of enrollment samples on identification accuracy and \(k\)-NN query latency on the LoRa dataset. Results are reported as mean \(\pm\) standard deviation.}
\label{tab:enrollment_size}
\begin{tabular}{ccc}
\toprule
\begin{tabular}[c]{@{}c@{}}Enrollment Samples\\per Device\end{tabular}
& ACC ($\uparrow$)
& \begin{tabular}[c]{@{}c@{}}Latency\\(ms/sample) ($\downarrow$)\end{tabular} \\
\midrule
100
& $0.483 \pm 0.078$
& $\mathbf{0.0570 \pm 0.0006}$ \\

200
& $0.522 \pm 0.077$
& $0.0698 \pm 0.0002$ \\

500
& $\mathbf{0.761 \pm 0.075}$
& $0.1015 \pm 0.0005$ \\
\bottomrule
\end{tabular}
\end{table}

\section{Conclusion}
\label{sec:Conclusion}
This work investigated high-openness RFFI under limited base-device class diversity, where a frozen feature extractor must support post-training device enrollment, enrolled-device identification, and unregistered-device rejection. To address this setting, HoRFFI incorporates similarity-aware compression, label-supervised discrimination, and inter-sample similarity regularization into the SVIB objective. The resulting embeddings transfer more effectively to devices absent during feature-extractor training and can be directly used with a \(k\)-NN-based enrollment and identification procedure. Experiments on the LoRa and Wi-Fi datasets showed consistent gains in newly enrolled-device identification across the evaluated openness levels, together with competitive or superior unregistered-device
rejection performance.


\ifanonymous
\else
\section*{Acknowledgment}
This work was supported by [funding agency and grant number].
\fi


\bibliographystyle{IEEEtran}
\bibliography{HoRFFI_references}

@article{aouedi2024survey,
  title={A survey on intelligent Internet of Things: Applications, security, privacy, and future directions},
  author={Aouedi, Ons and Vu, Thai-Hoc and Sacco, Alessio and Nguyen, Dinh C and Piamrat, Kandaraj and Marchetto, Guido and Pham, Quoc-Viet},
  journal={IEEE Commun. Surveys Tuts.},
  volume={27},
  number={2},
  pages={1238--1292},
  year={2025},
  }

@article{liu2021machine,
  title={Machine learning for the detection and identification of Internet of Things devices: A survey},
  author={Liu, Yongxin and Wang, Jian and Li, Jianqiang and Niu, Shuteng and Song, Houbing},
  journal={IEEE Internet Things J.},
  volume={9},
  number={1},
  pages={298--320},
  year={2022},
  }

@article{zhang2021radio,
  title={Radio frequency fingerprint identification for narrowband systems, modelling and classification},
  author={Zhang, Junqing and Woods, Roger and Sandell, Magnus and Valkama, Mikko and Marshall, Alan and Cavallaro, Joseph},
  journal={IEEE Trans. Inf. Forensics Security},
  volume={16},
  pages={3974--3987},
  year={2021},
  }

@article{cai2024toward,
  title={Toward intelligent lightweight and efficient UAV identification with RF fingerprinting},
  author={Cai, Zhenxin and Wang, Yu and Jiang, Qi and Gui, Guan and Sha, Jin},
  journal={IEEE Internet Things J.},
  volume={11},
  number={15},
  pages={26329--26339},
  year={2024},
  }

@article{xie2021generalizable,
  title={A generalizable model-and-data driven approach for open-set RFF authentication},
  author={Xie, Renjie and Xu, Wei and Chen, Yanzhi and Yu, Jiabao and Hu, Aiqun and Ng, Derrick Wing Kwan and Swindlehurst, A Lee},
  journal={IEEE Trans. Inf. Forensics Security},
  volume={16},
  pages={4435--4450},
  year={2021},
  }

@inproceedings{cai2025open,
  title={Open set RF fingerprinting identification: A joint prediction and Siamese comparison framework},
  author={Cai, Donghong and Shan, Jiahao and Gao, Ning and He, Bingtao and Chen, Yingyang and Jin, Shi and Fan, Pingzhi},
  booktitle={Proc. IEEE Int. Conf. Commun. (ICC)},
  pages={1007--1012},
  year={2025},
  month=jun,
  address={Montreal, QC, Canada},
  }

@article{yu2019robust,
  title={A robust RF fingerprinting approach using multisampling convolutional neural network},
  author={Yu, Jiabao and Hu, Aiqun and Li, Guyue and Peng, Linning},
  journal={IEEE Internet Things J.},
  volume={6},
  number={4},
  pages={6786--6799},
  year={2019},
  }

@inproceedings{alemi2016deep,
  title={Deep variational information bottleneck},
  author={Alemi, Alexander A. and Fischer, Ian and Dillon, Joshua V. and Murphy, Kevin},
  booktitle={Proc. Int. Conf. Learn. Represent. (ICLR)},
  year={2017},
  month=apr,
  address={Toulon, France},
}

@article{shen2022towards,
  title={Towards scalable and channel-robust radio frequency fingerprint identification for LoRa},
  author={Shen, Guanxiong and Zhang, Junqing and Marshall, Alan and Cavallaro, Joseph R},
  journal={IEEE Trans. Inf. Forensics Security},
  volume={17},
  pages={774--787},
  year={2022},
  }

@article{peng2024channel,
  title={Channel-robust radio frequency fingerprint identification for cellular uplink LTE devices},
  author={Peng, Linning and Peng, Haichuan and Fu, Hua and Liu, Ming},
  journal={IEEE Internet Things J.},
  volume={11},
  number={10},
  pages={17154--17169},
  year={2024},
  }

@article{jiang2024radio,
  title={Radio frequency fingerprint identification using conditional generative adversarial network for SAT-AIS},
  author={Jiang, Qi and Sha, Jin},
  journal={IEEE Trans. Aerosp. Electron. Syst.},
  volume={61},
  number={1},
  pages={593--602},
  year={2025},
  }

@article{shen2021radio,
  title={Radio frequency fingerprint identification for LoRa using deep learning},
  author={Shen, Guanxiong and Zhang, Junqing and Marshall, Alan and Peng, Linning and Wang, Xianbin},
  journal={IEEE J. Sel. Areas Commun.},
  volume={39},
  number={8},
  pages={2604--2616},
  year={2021},
  }

@article{geng2020recent,
  title={Recent advances in open set recognition: A survey},
  author={Geng, Chuanxing and Huang, Sheng-jun and Chen, Songcan},
  journal={IEEE Trans. Pattern Anal. Mach. Intell.},
  volume={43},
  number={10},
  pages={3614--3631},
  year={2021},
  }

@inproceedings{bendale2015towards,
  title={Towards open world recognition},
  author={Bendale, Abhijit and Boult, Terrance},
  booktitle={Proc. IEEE Conf. Comput. Vis. Pattern Recognit. (CVPR)},
  pages={1893--1902},
  year={2015},
  month=jun,
  address={Boston, MA, USA},
}

@article{huang2024radio,
  title={Radio frequency fingerprint extraction and authentication towards open set in noisy channels},
  author={Huang, Renhui and Peng, Xinyong and Chai, Zhi and Li, Mingye and Ren, Jiawei and Yang, Xuelin},
  journal={Digit. Signal Process.},
  volume={146},
  pages={104363},
  year={2024},
  publisher={Elsevier},
  doi={10.1016/j.dsp.2023.104363}
}

@inproceedings{han2025open,
  title={Open-world radio frequency fingerprint identification via augmented semi-supervised learning},
  author={Han, Zehua and Xiao, Jing and Zhao, Qirui and Cui, Zhexuan and Wang, Yufeng and Zhang, Duona and Ding, Wenrui},
  booktitle={Proc. AAAI Conf. Artif. Intell.},
  volume={39},
  number={1},
  pages={264--272},
  year={2025},
  month={Feb.--Mar.},
  address={Philadelphia, PA, USA},
}

@inproceedings{cao2021open,
  title={Open-world semi-supervised learning},
  author={Cao, Kaidi and Brbi{\'c}, Maria and Leskovec, Jure},
  booktitle={Proc. Int. Conf. Learn. Represent. (ICLR)},
  year={2022},
  month=apr,
  address={Virtual Conf.},
}

@article{liu2023openworld,
      title={Open-world Semi-supervised Novel Class Discovery}, 
      author={Jiaming Liu and Yangqiming Wang and Tongze Zhang and Yulu Fan and Qinli Yang and Junming Shao},
      year={2023},
      journal={arXiv preprint arXiv:2305.13095}

}

@inproceedings{tishby2000information,
  title={The information bottleneck method},
  author={Tishby, Naftali and Pereira, Fernando C. and Bialek, William},
  booktitle={Proc. 37th Annu. Allerton Conf. Commun., Control, Comput.},
  pages={368--377},
  year={1999},
  month=sep,
  address={Monticello, IL, USA},
}

@inproceedings{cen2023enlarging,
  title={Enlarging instance-specific and class-specific information for open-set action recognition},
  author={Cen, Jun and Zhang, Shiwei and Wang, Xiang and Pei, Yixuan and Qing, Zhiwu and Zhang, Yingya and Chen, Qifeng},
  booktitle={Proc. IEEE/CVF Conf. Comput. Vis. Pattern Recognit. (CVPR)},
  pages={15295--15304},
  year={2023},
  month=jun,
  address={Vancouver, BC, Canada},
}

@article{scheirer2012toward,
  title={Toward open set recognition},
  author={Scheirer, Walter J and de Rezende Rocha, Anderson and Sapkota, Archana and Boult, Terrance E},
  journal={IEEE Trans. Pattern Anal. Mach. Intell.},
  volume={35},
  number={7},
  pages={1757--1772},
  year={2013},
  }

@inproceedings{kingma2014autoencoding,
  title={Auto-Encoding Variational Bayes},
  author={Kingma, Diederik P and Welling, Max},
  booktitle={Proc. 2nd Int. Conf. Learn. Represent. (ICLR)},
  year={2014},
  address={Banff, AB, Canada},
  month=apr,
  note={Conference Track Proceedings}
}

@inproceedings{wang2011information,
  title={Information theoretical clustering via semidefinite programming},
  author={Wang, Meihong and Sha, Fei},
  booktitle={Proc. 14th Int. Conf. Artif. Intell. Statist. (AISTATS)},
  series={Proc. Mach. Learn. Res.},
  volume={15},
  pages={761--769},
  year={2011},
  month=apr,
  address={Fort Lauderdale, FL, USA},
}

@inproceedings{sankhe2019oracle,
  title={ORACLE: Optimized radio classification through convolutional neural networks},
  author={Sankhe, Kunal and Belgiovine, Mauro and Zhou, Fan and Riyaz, Shamnaz and Ioannidis, Stratis and Chowdhury, Kaushik},
  booktitle={Proc. IEEE INFOCOM},
  pages={370--378},
  year={2019},
  month={Apr.--May},
  address={Paris, France},
  }

@article{zhao2025survey,
  title={A Survey of Physical Layer Authentication for Millimeter-Wave MIMO Systems},
  author={Zhao, Xu},
  journal={J. Netw. Netw. Appl.},
  volume={5},
  number={3},
  pages={137--147},
  year={2025},
  publisher={Institute of Electronics and Computer}
}

@article{zhang2025physical,
  title={Physical layer-based device fingerprinting for wireless security: From theory to practice},
  author={Zhang, Junqing and Ardizzon, Francesco and Piana, Mattia and Shen, Guanxiong and Tomasin, Stefano},
  journal={IEEE Trans. Inf. Forensics Security},
  volume={20},
  pages={5296--5325},
  year={2025},
  doi={10.1109/TIFS.2025.3570118}
}

@inproceedings{shen2021spectrogram,
  title={Radio frequency fingerprint identification for LoRa using spectrogram and CNN},
  author={Shen, Guanxiong and Zhang, Junqing and Marshall, Alan and Peng, Linning and Wang, Xianbin},
  booktitle={Proc. IEEE INFOCOM},
  pages={1--10},
  year={2021},
  month=may,
  address={Virtual Conf.},
  }

@inproceedings{xie2025towards,
  title={Towards robust RF fingerprint identification using spectral regrowth and carrier frequency offset},
  author={Xie, Lingnan and Peng, Linning and Zhang, Junqing},
  booktitle={Proc. IEEE INFOCOM},
  pages={1--10},
  year={2025},
  month=may,
  address={London, United Kingdom},
  }

@article{sun2026enhancing,
  title={Enhancing open-set RFF recognition with cGAN: Generating multiple unknown classes},
  author={Sun, Haohao and Zou, Cong and Wang, Qiexiang and Wang, Jian and Zhang, Xudong},
  journal={IEEE Internet Things J.},
  volume={13},
  number={6},
  pages={12388--12404},
  year={2026},
  doi={10.1109/JIOT.2026.3652146}
}

@article{chen2023extreme,
  title={An extreme value theory-based approach for reliable drone RF signal identification},
  author={Chen, Yufan and Zhu, Lei and Jiao, Yuchen and Yao, Changhua and Cheng, Kaixin and Gu, Yuantao},
  journal={IEEE Trans. Cogn. Commun. Netw.},
  volume={10},
  number={2},
  pages={454--469},
  year={2024},
  }

@article{xie2025novel,
  title={A novel radio frequency fingerprint identification scheme for few-shot open-set recognition},
  author={Xie, Wei and Wang, Hongjun and Shen, Zhexian and Li, Xinhao and Liu, Zhiquan and Jiang, Hao},
  journal={IEEE Internet Things J.},
  volume={12},
  number={13},
  pages={25691--25706},
  year={2025},
  doi={10.1109/JIOT.2025.3559183}
}

@article{wu2024open,
  title={Open set RF fingerprint identification for wireless communication devices},
  author={Wu, Chaopeng and Chen, Shiwen and Sun, Gangyin and Fang, Haikun},
  journal={IEEE Wireless Commun. Lett.},
  volume={14},
  number={3},
  pages={776--780},
  year={2025},
  publisher={IEEE},
  doi={10.1109/LWC.2024.3523271}
}

@article{yin2024multi,
  title={Multi-channel CNN-based open-set RF fingerprint identification for LTE devices},
  author={Yin, Pengcheng and Peng, Linning and Shen, Guanxiong and Zhang, Junqing and Liu, Ming and Fu, Hua and Hu, Aiqun and Wang, Xianbin},
  journal={IEEE Trans. Cogn. Commun. Netw.},
  volume={10},
  number={5},
  pages={1788--1800},
  year={2024},
  }

@article{sun2026knowledge,
  title={From Knowledge Graphs to Decision Boundaries: Separable Embeddings for Open-Set Specific Emitter Identification},
  author={Sun, Lu and Xue, Rui and Zha, Haoran and Cao, Shunyao and Li, Ruofei and Gui, Guan and Lin, Yun},
  journal={IEEE Trans. Veh. Technol.},
  year={2026},
  }

@article{li2023class,
  title={A class-incremental approach with self-training and prototype augmentation for specific emitter identification},
  author={Li, Dingzhao and Qi, Jie and Hong, Shaohua and Deng, Pengfei and Sun, Haixin},
  journal={IEEE Trans. Inf. Forensics Security},
  volume={19},
  pages={1714--1727},
  year={2024},
  }

@inproceedings{hadsell2006dimensionality,
  title={Dimensionality reduction by learning an invariant mapping},
  author={Hadsell, Raia and Chopra, Sumit and LeCun, Yann},
  booktitle={Proc. IEEE Comput. Soc. Conf. Comput. Vis. Pattern Recognit. (CVPR)},
  volume={2},
  pages={1735--1742},
  year={2006},
  month=jun,
  address={New York, NY, USA},
  }

@inproceedings{khosla2020supervised,
  title={Supervised contrastive learning},
  author={Khosla, Prannay and Teterwak, Piotr and Wang, Chen and Sarna, Aaron and Tian, Yonglong and Isola, Phillip and Maschinot, Aaron and Liu, Ce and Krishnan, Dilip},
  booktitle={Adv. Neural Inf. Process. Syst.},
  volume={33},
  pages={18661--18673},
  year={2020},
  month=dec,
  address={Virtual Conf.},
}

@article{shen2025class,
  title={A class incremental learning method with forward-compatible and covariance-aware for specific emitter identification},
  author={Shen, Xiaoyu and Zhang, Jiang and Qiao, Xiaoqiang and Shang, Zhihui and Wang, Min and Zhang, Tao},
  journal={IEEE Commun. Lett.},
  volume={29},
  number={9},
  pages={2148--2152},
  year={2025},
  doi={10.1109/LCOMM.2025.3588238}
}

@inproceedings{schroff2015facenet,
  title={Facenet: A unified embedding for face recognition and clustering},
  author={Schroff, Florian and Kalenichenko, Dmitry and Philbin, James},
  booktitle={Proc. IEEE Conf. Comput. Vis. Pattern Recognit. (CVPR)},
  pages={815--823},
  year={2015},
  month=jun,
  address={Boston, MA, USA},
}

@inproceedings{sohn2016improved,
  title={Improved deep metric learning with multi-class N-pair loss objective},
  author={Sohn, Kihyuk},
  booktitle={Adv. Neural Inf. Process. Syst.},
  volume={29},
  year={2016},
  month=dec,
  address={Barcelona, Spain},
}

@inproceedings{wen2016discriminative,
  title={A discriminative feature learning approach for deep face recognition},
  author={Wen, Yandong and Zhang, Kaipeng and Li, Zhifeng and Qiao, Yu},
  booktitle={Proc. Eur. Conf. Comput. Vis. (ECCV)},
  pages={499--515},
  year={2016},
  month=oct,
  address={Amsterdam, The Netherlands},
  organization={Springer}
}

@article{jagannath2022comprehensive,
  title={A comprehensive survey on radio frequency (RF) fingerprinting: Traditional approaches, deep learning, and open challenges},
  author={Jagannath, Anu and Jagannath, Jithin and Kumar, Prem Sagar Pattanshetty Vasanth},
  journal={Comput. Netw.},
  volume={219},
  pages={109455},
  year={2022},
  publisher={Elsevier}
}

@inproceedings{musgrave2020metric,
  title={A metric learning reality check},
  author={Musgrave, Kevin and Belongie, Serge and Lim, Ser-Nam},
  booktitle={Proc. Eur. Conf. Comput. Vis. (ECCV)},
  pages={681--699},
  year={2020},
  month=aug,
  address={Glasgow, United Kingdom},
  organization={Springer}
}

@article{shen2026towards,
  title={Towards Efficient Deep Learning in RF Fingerprint Identification with OverlapConv},
  author={Shen, Yuxiang and Zeng, Shuiguang and Tan, Zhiyuan and Shen, Yulong and Zhao, Dongmei and Song, Houbing Herbert},
  journal={IEEE Internet Things J.},
  volume={13},
  number={10},
  pages={22514--22532},
  year={2026},
  publisher={IEEE}
}

@article{wang2025review,
  title={A Review of Crystal Oscillators Imperfection: Linking Frequency Deviations to Carrier Frequency Offset and Phase Noise},
  author={Wang, Jianing and Liu, Xin},
  journal={J. Netw. Netw. Appl.},
  volume={4},
  number={4},
  pages={165--171},
  year={2025},
  publisher={Institute of Electronics and Computer}
}

\end{document}